\documentstyle[scibib,preprint,epsfig,aps]{revtex}

\begin{document}

\draft
\title{The Spin Excitation Spectrum in Superconducting 
${\bf YBa_2Cu_3O_{6.85}}$}
\author{P. Bourges$^1$, Y. Sidis$^1$, H.F. Fong$^2$, L.P. Regnault$^3$, J. Bossy$^4$, A. Ivanov$^5$, B.~Keimer$^{2,6}$}

\address{1 - Laboratoire L\'eon Brillouin, CEA-CNRS, CE Saclay, 91191 Gif 
sur Yvette, France}
\address{2 - Department of Physics, Princeton University, Princeton, NJ 
08544, USA}
\address{3 - CEA Grenoble, D\'epartement de Recherche Fondamentale sur
la mati\`ere Condens\'ee, 38054~Grenoble cedex~9, France}
\address{4 - CNRS-CRTBT, BP 166, 38042 Grenoble cedex~9, France}
\address{5 - Institut Laue-Langevin, 156X, 38042 Grenoble Cedex 9, France}
\address{6 - Max-Planck-Institut f\"ur Festk\"orperforschung,
70569 Stuttgart, Germany}

\maketitle

\vskip 0.5 cm 
\centerline{Science, {\bf 288} p 1234 (2000).}
 
\clearpage

A comprehensive inelastic neutron scattering study of
magnetic excitations
in the near-optimally doped 
high-temperature superconductor ${\rm
YBa_{2}Cu_{3}O_{6.85}}$ is presented.  The spin correlations in the 
normal state are commensurate with the crystal lattice, and the intensity 
is peaked around the wave vector characterizing the antiferromagnetic state 
of the insulating precursor ${\rm YBa_{2}Cu_{3}O_{6}}$. Profound 
modifications of
the spin excitation spectrum appear abruptly below the superconducting
transition temperature,
$\rm T_C$, where a commensurate resonant mode and a set of weaker
incommensurate
peaks develop. The data are consistent with models based on an underlying
two-dimensional 
Fermi surface that predict a continuous, downward
dispersion relation
connecting the resonant mode and the incommensurate
excitations. 
The magnetic incommensurability in the ${\rm
YBa_{2}Cu_{3}O_{6+x}}$ system 
is thus not simply related to that of
another high temperature 
superconductor, ${\rm La_{2-x}Sr_xCuO_{4}}$,
where incommensurate peaks persist 
well above $\rm T_C$. The temperature
dependent incommensurability is difficult to
reconcile with interpretations based on charge stripe formation in 
${\rm YBa_{2}Cu_{3}O_{6+x}}$ near optimum doping.


\clearpage 

Electronic conduction in the high temperature superconductors cuprate takes
place predominantly in ${\rm CuO_2}$ layers. Most theories therefore regard
the electronic state that forms the basis of high temperature
superconductivity as an essentially two-dimensional (2D) strongly correlated
metal. The ${\rm CuO_2}$ sheets in one family of copper oxides (${\rm %
La_{2-x}Sr_xCuO_{4}}$) have, however, been shown to be unstable against the
formation of 1D ``charge stripes"\cite{tranquada} even near doping levels
where the superconducting transition temperature ${\rm T_c}$ is maximum.
This observation has boosted models in which the underlying electronic
instability is one-dimensional and the formation of (static or fluctuating)
stripes is an essential precondition for high temperature superconductivity
\cite{stripes}. However, ${\rm La_{2-x}Sr_xCuO_{4}}$ has some low energy
phonon modes conducive to stripe formation that are not generic to the 
high-${\rm T_C}$ compounds, and the maximum ${\rm T_C}$ in this system 
is anomalously low. It is therefore important to test whether stripe 
based scenarios are viable in other cuprates with higher ${\rm T_C}$,
where this lattice dynamical peculiarity is not present.

The most salient signature of charge stripes is an associated (static or
dynamic) spin density modulation that can be detected by neutron scattering.
In ${\rm La_{2- x}Sr_xCuO_{4}}$ this modulation manifests itself as four
well-defined incommensurate peaks at $Q_{\delta}=(\pi(1\pm\delta),\pi)$ and 
$(\pi,\pi(1\pm\delta))$ (in square lattice notation with unit
lattice constant) in the magnetic spectrum, which are interpreted as arising
from two 1D domains \cite{mason,thurston,sylv,yamada}. Neutron scattering
experiments on the ${\rm YBa_{2}Cu_{3}O_{6+x}}$ system have, however,
revealed excitations that are peaked at $Q_{AF}=(\pi,\pi)$ 
\cite{rossat91,mook93,fong95,bourges96,fong96,revue-cargese,revue-lpr}
, the ordering wave vector of the 2D antiferromagnetic state observed when
the doping level is reduced to zero. In particular, the commensurate
``resonance peak" at $Q=(\pi,\pi)$ that dominates the spectrum in the
superconducting state
\cite{fong95,bourges96,fong96,revue-cargese,revue-lpr,tony97,epl},
 is difficult to reconcile with scenarios based on
fluctuating 1D domains incommensurate with the host lattice. Recently, an
incommensurate pattern with a four-fold symmetry reminiscent of 
${\rm La_{2-x}Sr_xCuO_{4}}$ has also been discovered in some constant-energy cuts
of the magnetic spectrum of underdoped ${\rm YBa_{2}Cu_{3}O_{6.6}}$\cite
{incdai,mookinc,arai}, which was taken as experimental support for
stripe-based scenarios of superconductivity. We report a neutron
scattering study of near-optimally doped ${\rm YBa_{2}Cu_{3}O_{6.85}}$ ($%
{\rm T_c = 89}$K) demonstrating that (unlike in ${\rm La_{2-x}Sr_xCuO_{4}}$)
the incommensurate pattern appears only below ${\rm T_c}$. Magnetic
excitations in the normal state are commensurate and centered at $Q=(\pi,\pi)
$. Our data are consistent with 2D Fermi liquid-like theories (not invoking stripes) \cite{vdM,OP,BL,levin,VooandWu}, and especially that
which predicts a continuous, downward dispersion of the magnetic 
resonance peak \cite{OP}. 

The experiments have been performed on a large twinned single crystal 
(mass $\sim $
9.5 g) grown using the top seed melt texturing method \cite{long-tony}. The
sample was subsequently annealed in oxygen and displays a sharp
superconducting transition ($T_C$) at 89 K measured by a neutron de-polarization
technique that is sensitive to the entire bulk \cite{long-tony}. 
Experiments were carried out on two triple axis
spectrometers: IN8 at the Institut Laue Langevin, Grenoble (France), and 2T
at the Laboratoire L\'{e}on Brillouin, Saclay (France)\cite{expdetails}.
 Two different scattering geometries were
used on both spectrometers. On IN8, the (130) and (001) reciprocal
directions were within the horizontal scattering plane. [We quote the
wavevector {\bf Q}=(H,K,L) in units of the tetragonal reciprocal lattice
vectors $2\pi /a=2\pi /b=1.63$ \AA $^{-1}$ and $2\pi /c=0.53$ \AA $^{-1}$].
On 2T, an unconventional scattering geometry has been employed with the
(100) and (011) reciprocal directions spanning the scattering plane. In both
scattering geometries in-plane wavevectors equivalent to $\equiv (\pi ,\pi )$
can be reached, with an out-of-plane wave vector component close to the
maximum of the structure factor of low energy excitations \cite{long-tony}.
In addition, wavevectors of the form ${\bf Q}=(H,K,1.7)$ around the
antiferromagnetic wavevector were accessible on 2T by controlling the tilt
angle, allowing a 2D mapping of the neutron intensity in the tetragonal
basal momentum space and for a fixed energy transfer.

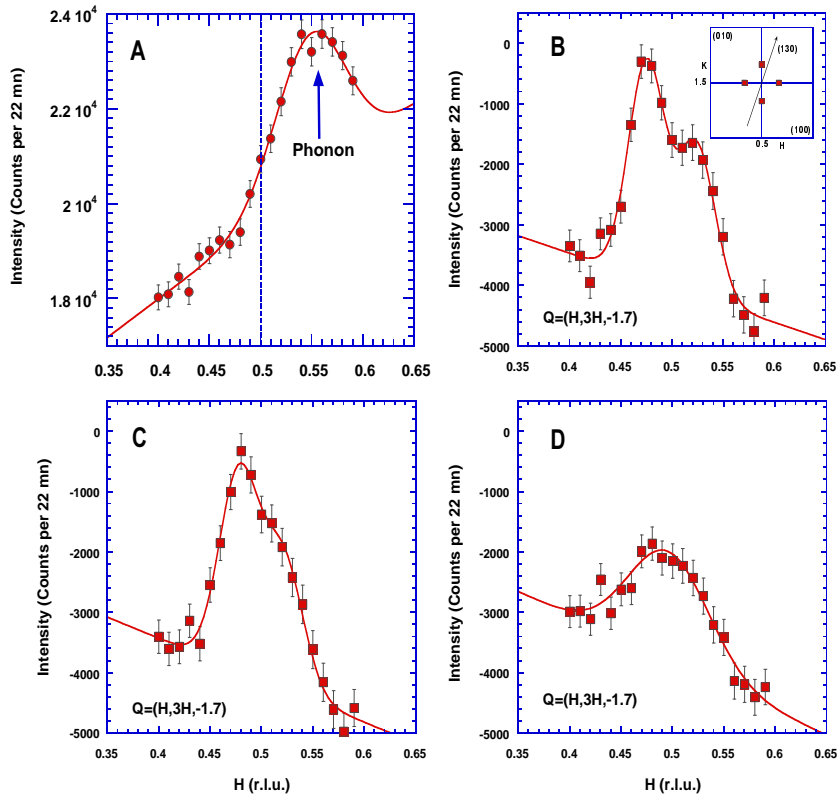
\begin{figure}[tbp]
\caption{ ({\bf A}) Raw (uncorrected) room temperature scan which exhibits a phonon
peak around $H=0.57$. This scan displays none of the features of panels
({\bf B-D}). The inset in ({\bf B})  is a sketch of the reciprocal space around the AF wavevector.
The squares represent the locus of maximum magnetic intensity in the superconducting state. The arrow
the represents the (130) direction of the scans. ({\bf B-D}) Neutron intensity 
of constant-energy scans at $E=35$ meV performed along the (130) 
direction, with the room temperature
scan subtracted: ({\bf B}) I(11K)-I(300K) ({\bf C}) I(70K)-I(300K) 
 ({\bf D}) I(100K)-I(300K). The momentum resolution was 0.03 r.l.u. 
(=0.15 \AA$^{-1}$) along (1,3,0) direction and 0.3 \AA$^{-1}$ along the 
vertical direction. The energy
resolution was 5 meV. We cannot rule out a small AF intensity ($\sim$ 4
times less than at 100 K) at room temperature.}
\label{qscans}
\end{figure}

The magnetic resonance peak is observed at $E_{r}=41$ meV and $Q=Q_{AF}$ in
our sample \cite{long-tony} in agreement with previous reports for similar
oxygen content \cite{rossat91,revue-cargese}. Here we present data obtained
at energies below the resonance where the magnetic signal is significantly
weaker. Using established procedures, we use the temperature dependence of
the magnetic intensity (which strongly decreases upon heating) to separate
it from the phonon scattering that gradually increases with increasing
temperature (Fig. \ref{qscans}A). Figs. \ref{qscans}B-D show constant-energy
scans at E=35 meV performed along the (130) direction (arrow in the inset of
Fig. \ref{qscans}B). The phonon contribution has been subtracted for
clarity. At low temperature, deep in the superconducting state, the magnetic
scattering exhibits a double peak structure along $(H,3H,0)$. At 70 K, the
intensity is still peaked at incommensurate wavevectors, but the
discommensuration is slightly reduced. A single broad feature peaked at $H$%
=0.5 is observed at 100 K in the normal state.

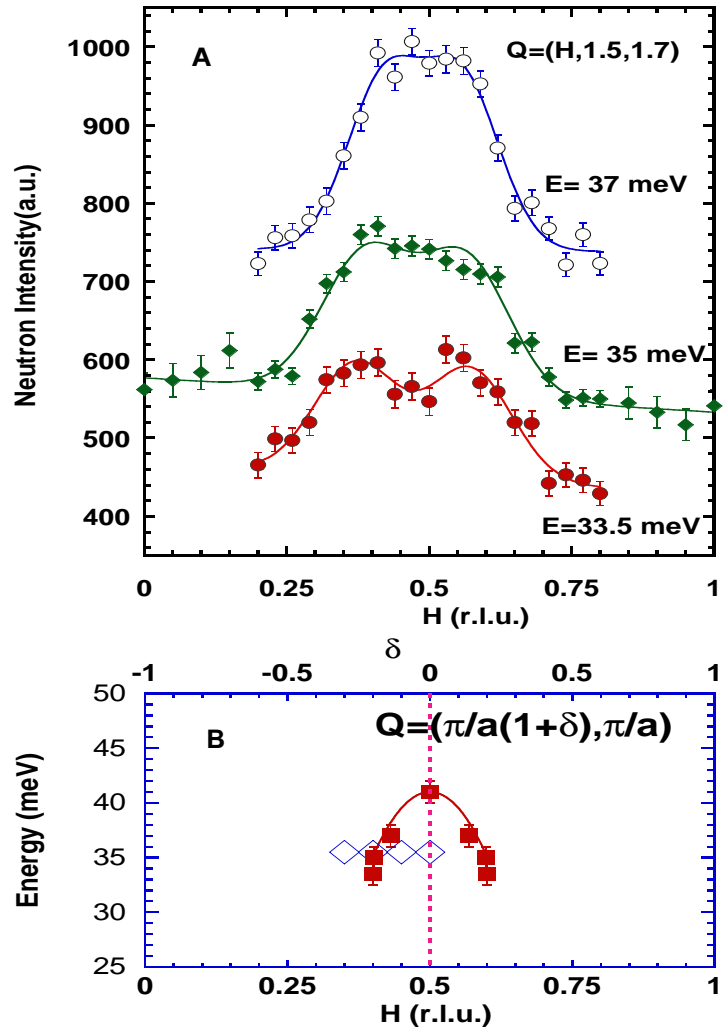
\begin{figure}[tbp]
\caption[[bla]{ ({\bf A}) Constant-energy scans performed along the $H$ direction. 
The scans are offset by 120 counts from one another. The momentum 
resolution (FWHM) was 0.14 r.l.u. along $H$ and 0.1 r.l.u. along $K$. 
The energy resolution was 4 meV. The lines are Gaussian
displaced by $\pm\protect\delta H$ from $Q_{AF}$. The momentum
transfer along $c^*$ was fixed to the maximum of the magnetic structure
factor, $L=1.7$. The phonon background measured at room temperature has been
subtracted from the data after proper correction for the Bose population factor
\cite{epl}. ({\bf B})  Dispersion
of the incommensurate peaks observed in YBCO$_{6.85}$, deduced from (a) as described in the text. The square
at $H=0.5$ represents the resonance peak, $E_r$=41 meV. The other squares
at $0.5 \pm \Delta H$ come from the Gaussian fits shown in the lower panel.
The open diamonds indicate the four wavevectors at E=35 meV where the
temperature dependence has been studied (see Fig. \ref{temperature}). }
\label{w-scans}
\end{figure}

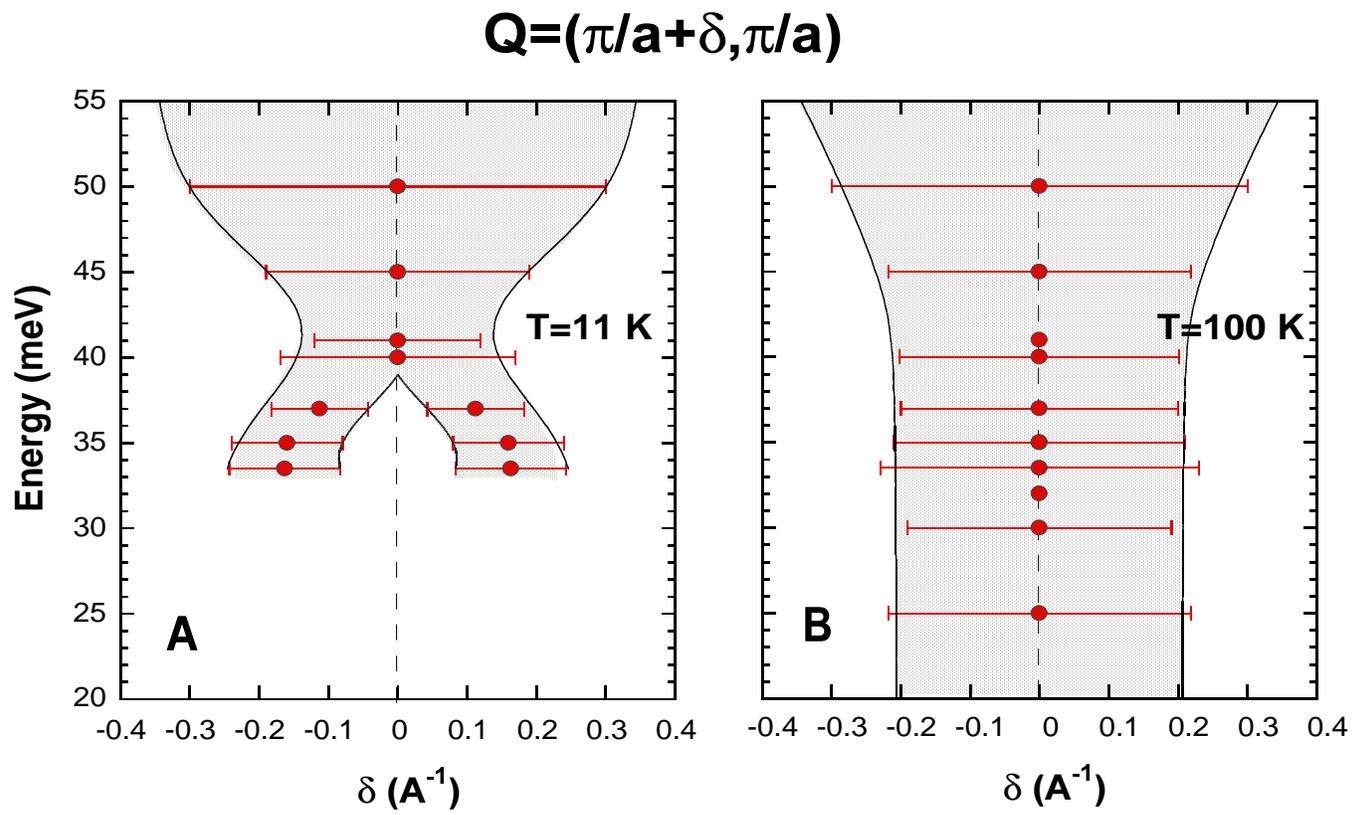
\begin{figure}[tbp]
\caption[[bla]{ Overall momentum dependence of the magnetic response in YBCO$%
_{6.85}$ obtained ({\bf A}) in the superconducting state at T=11 K  and
({\bf B}) in the normal state at T=100 K. }
\label{momentum}
\end{figure}

We have performed a comprehensive set of measurements in order to chart out
in detail the spectral rearrangement indicated in Fig. \ref{qscans}. As an
example, typical constant-energy scans were taken at low temperature 
along the $H$ direction with $K$=1.5 (Fig. \ref{w-scans}A). The magnetic
intensity is maximum at incommensurate wavevectors displaced from $(0.5,1.5)$
along $H$, and the incommensurability in this direction depends continuously
on energy and forms a dispersing branch that closes at $E_{r}$.
Accordingly, the maximum of the spin susceptibility at wave 
vector sufficiently far from ($\pi,\pi$) is shifted to lower energy 
than  the resonant peak energy. The profile
shapes are influenced by the instrumental resolution, and a deconvolution is
required to accurately extract the peak positions. We found that a good global 
fit to the peak positions in both scattering geometries
could be obtained by a convolution of the spectrometer resolution with the
dispersion relation $E=\sqrt{E_{r}^{2}-(\alpha q)^{2}}$ with $\alpha
=(125\pm 15)$ meV- \AA\cite{isotrope}. This downward dispersion 
(Fig. \ref{w-scans}B) is shown along with the fitted peak positions. 
A neutron intensity map in the $(H,K)$ reciprocal lattice 
plane measured at E=35 meV and T=12 K (not shown here)
indicates that the intensity is not uniform along the dispersion 
surface. As in more 
underdoped samples \cite{mookinc,arai}, the intensity pattern has fourfold 
symmetry, with maxima shown as full squares in the inset of 
Fig. \ref{qscans}B. Although the data have not been analyzed in this manner, 
the overall momentum dependence of the spin
susceptibility reported in Fig. \ref{momentum} below ${\rm T_{c}}$ is
consistent with that reported in Refs. \cite{incdai,arai} as well as that in
Refs. \cite{bourges96,epl} where the incommensurability was not resolved.

Similar scans were repeated in the normal state (Fig. \ref{momentum}B) 
where the data thus far reported are inconclusive about the 
incommensurability. For all energies reported in Fig. \ref{w-scans}, the 
normal state response can be systematically fitted by a
single broad line (as shown at E= 35 meV in Fig. \ref{qscans}D). Fig. \ref
{momentum} shows that the momentum width at T=100 K is energy independent up
to 45 meV where the peaks begin to broaden \cite{highenergy}. The $q$-width
at low energy, $\Delta_q=$ 0.21 \AA$^{-1}$ (HWHM) after deconvolution,
agrees with previous reports for a similar oxygen content\cite{epl,balatsky}.
Below ${\rm T_c}$, the spin dynamics exhibit a more complex momentum
dependence (Fig. \ref{momentum}A). While the momentum shape at energies above
45 meV is largely unaffected by superconductivity, there is a large increase
in intensity and narrowing of the $q$-width at $E_r = 41$ meV which is
accompanied by a much weaker incommensurate response in a narrow energy
range below $E_r$. Finally, the intensity is strongly reduced below 30 meV,
which could result from the opening of a "spin-gap" below ${\rm T_c}$ as
suggested by previous measurements \cite{rossat91,revue-cargese}.

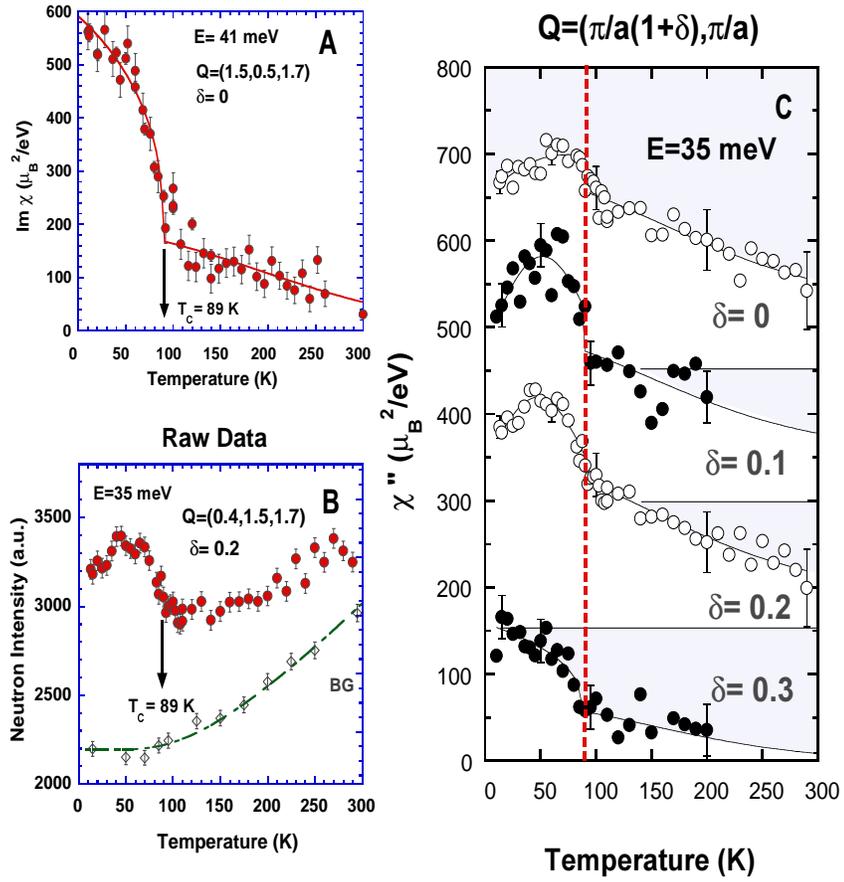
\begin{figure}[tbp]
\caption{({\bf A})  Temperature dependence of spin susceptibility in absolute units
at the resonance energy $E_r$= 41 meV. ({\bf B})  Temperature dependence of the
neutron intensity at E= 35 meV and at the incommensurate wavevector $Q_{%
\protect\delta}=(1.5,0.4,1.7)$ (full circles). The open squares represent
the background, determined by constant-energy scans (equivalent to the one
shown in Fig. \ref{w-scans}A). ({\bf C})  Temperature dependence of spin
susceptibility in absolute units at $E$= 35 meV for 4 momentum transfers
along $a^*$ (4 open diamonds in Fig. \ref{w-scans}B). The
curves have been shifted by 150 $\protect\mu_B^2$/eV from one another. The
susceptibility has been obtained by background subtraction and correction 
for the temperature factor $1/(1-\exp{(-E/k_B T}))$.
A marked change at $\rm T_c$ is observed at all wave vectors.}
\label{temperature}
\end{figure}

The incommensurate response thus appears to be intimately linked to the
occurrence of the resonance peak below ${\rm T_c}$ \cite
{rossat91,fong95,bourges96,fong96,revue-cargese,revue-lpr,tony97,long-tony,miami}%
. In order to complete this picture, we now describe an accurate
determination of the onset temperature of the incommensurate response and
compare it to that of the resonance peak \cite{mook93,bourges96,fong96}. The
temperature dependence of the resonance intensity is reproduced for the
present sample (Fig. \ref{temperature}A), and the raw neutron intensity 
(say without the phonon background subtracted) at the incommensurate
position $Q=(0.4,1.5,1.7)$ and E=35 meV is shown versus temperature 
(Fig. \ref{temperature}B). Processed data at E=35 meV (phonons subtracted 
and converted to absolute units \cite{fong96,long-tony}) are 
shown (Fig. \ref{temperature}C) a series of wavevectors 
(sketched by the open diamonds in Fig. \ref{w-scans}B) spanning the 
locus of maximum
intensity at low temperatures. Remarkably, all curves show a precipitous
upturn at ${\rm T_c}$, demonstrating clearly that both the resonance peak
and the incommensurability are induced by superconductivity. An indication
of similar behavior had already been found in ${\rm YBa_2 Cu_3 O_{6.7}}$ ($%
{\rm T_c}$=67 K) \cite{miami} so that this behavior seems to be generic to
the ${\rm YBa_2 Cu_3 O_{6+x}}$ superconductor.

Fig. \ref{temperature}C is also revealing in another respect. While all
curves show the same sharp initial upturn below ${\rm T_c}$, $\chi"({\bf q}%
,35\;{\rm meV})$ generally goes through a maximum at a temperature ${\rm T_m
({\bf q})}$ that increases as ${\bf q} \rightarrow (\pi,\pi)$. A monotonic
temperature dependence (${\rm T_m \rightarrow 0}$) like the one of the
resonance peak is seen only for {\bf q} close to the low temperature
incommensurate wave vector determined above. However, the difference $\chi"(%
{\rm T_m})-\chi"({\rm T_c})$ is identical within the errors for the three 
{\bf q}-points away from $(\pi,\pi)$. This dramatic behavior can be
straightforwardly understood as a consequence of the temperature (Fig. 1)
and energy (Figs. 2 and 3) dependent incommensurability already indicated
above. The resonance peak, together with the dispersion, forms quickly upon
cooling below ${\rm T_c}$ and then sweeps sequentially through the wave
vectors monitored in Figs. \ref{temperature}C.

In ${\rm YBa_{2}Cu_{3}O_{6.85}}$ the incommensurate response
is part of a continuous dispersion below the resonance peak that is strongly
renormalized upon approaching ${\rm T_{c}}$ and vanishes in the normal
state. This is in stark contrast to the behavior reported for ${\rm %
La_{2-x}Sr_{x}CuO_{4}}$ \cite{mason,thurston,sylv,yamada} where no change of
the peak position occurs across the superconducting temperature \cite{mason}%
. Only around room temperature does the incommensurate structure begin to
disappear \cite{aeppli}. Further, ${\delta }$ is energy-independent in ${\rm %
La_{2-x}Sr_{x}CuO_{4}}$, but depends strongly on doping over a wide range of
the phase diagram \cite{yamada}. Because of the energy dependence of ${%
\delta }$ discussed above and the small number of samples investigated thus
far, information about its doping dependence in ${\rm YBa_{2}Cu_{3}O_{6+x}}$
is still very incomplete. We note, however, that the discommensuration at $%
E=35$ meV ($=E_{r}-6$ meV) we found in ${\rm YBa_{2}Cu_{3}O_{6.85}}$, $%
\delta =0.10$ r.l.u. $\equiv 0.16$ \AA $^{-1}$, is equal to that reported in 
${\rm YBa_{2}Cu_{3}O_{6.6}}$ \cite{mookinc} at $E=24.5$ meV 
($=E_{r}{\acute{}}-9.5$ meV with $E_{r}{\acute{}}=34$ meV) within the 
experimental error. Finally, the incommensurate
fluctuations are only observed in a narrow energy range for fixed doping and
are further substantially weakened in fully oxidized ${\rm YBa_{2}Cu_{3}O_{7}%
}$\cite{fong95,bourges96,revue-cargese}.

In the light of these observations, the analogy between the incommensurate
spin excitations in ${\rm La_{2-x}Sr_{x}CuO_{4}}$ and ${\rm %
YBa_{2}Cu_{3}O_{6+x}}$ should not be overstated. Specifically, the
interpretation in terms of stripe-domain fluctuations with a well-defined,
doping dependent average periodicity, which is compelling and well
documented in ${\rm La_{2-x}Sr_{x}CuO_{4}}$, seems untenable in 
${\rm YBa_{2}Cu_{3}O_{6+x}}$ near optimal doping. Our data indicate that the
incommensurate excitations are continuously connected to the commensurate
resonance peak by a dispersion relation with a negative curvature
(Fig. \ref{qscans}B). The temperature dependence of Fig. \ref
{temperature} strongly supports a common origin of both phenomena. Some
aspects of the behavior we observed have, in fact, been anticipated in the
framework of microscopic models where the resonance peak is interpreted as 
a collective mode pulled below the gapped particle-hole spin-flip 
continuum\cite{vdM,OP,BL}. In particular, a downward 
dispersion has been predicted to arise naturally as a result of a 
momentum dependent pole in the spin susceptibility due to
antiferromagnetic interactions \cite{OP}. 
This picture also accounts for the rapidly diminishing intensity of 
the magnetic peaks away from $(\pi,\pi)$, as collective modes commonly lose 
oscillator strength upon approaching the continuum. Models
based on dynamical nesting induced by a modification of the band dispersions
\cite{BL,levin,VooandWu} may also be consistent with our data. The models favored by
our experimental results on near-optimally doped ${\rm YBa_{2}Cu_{3}O_{6+x}}$
are based on an interplay between band dispersions, Coulomb interactions
and the $d$-wave gap function in a two-dimensional correlated
electronic state.


\clearpage

\clearpage
Fig 1:

\epsfig{file=bourges_fig1.epsi,height=13 cm,width=13 cm} \clearpage
Fig 2:

\epsfig{file=bourges_fig2.epsi,height=15 cm,width=12 cm} \clearpage
Fig 3:

\vspace {1 cm } \epsfig{file=bourges_fig3.epsi,height=10 cm,width=13 cm} 
\clearpage
Fig 4:

\vskip 1 cm \ \ \
\epsfig{file=bourges_fig4.epsi,height=15 cm,width=12 cm}


\begin{thebibliography}{99}
\bibitem{expdetails}  On IN8, a Cu(111)
monochromator and a pyrolytic graphite PG(002) analyzer, set at a fixed
final energy of 35 meV, have been used. On 2T, PG(002) was employed as both
monochromator and analyzer with a 14.7 meV fixed final energy. In both
experiments, a PG filter was inserted into the scattered beam in order to
remove higher order contamination. The sample was
wrapped in aluminum foil and attached to the cold finger of a 
closed cycle helium refrigerator.
\bibitem{isotrope}  For a sake of simplicity, we describe the data with an
isotropic dispersion relation ($\alpha $ independent of $q$). A
slight anisotropy of $\alpha $ between the (100) and (110) directions, with an
associated intensity modulation, would
reproduce the detailed momentum dependence reported in \cite{mookinc}.
\end{thebibliography}
\end{document}